\documentclass[10pt, conference, compsocconf]{IEEEtran}

\let\labelindent\relax
\usepackage{amsmath}
\usepackage{listings}
\usepackage{array}
\usepackage{enumitem}
\usepackage{algorithm}
\usepackage{algpseudocode}
\usepackage{algorithmicx}
\usepackage{subfig}
\usepackage[pdftex]{graphicx}

\floatstyle{plain}
\newfloat{myalgo}{tbhp}{mya}

\begin{document}

\title{Achieving Speedup in Aggregate Risk Analysis using Multiple GPUs}

\author{\IEEEauthorblockN{A. K. Bahl\textsuperscript{1}, O. Baltzer\textsuperscript{2}, A. Rau-Chaplin\textsuperscript{2}, B. Varghese\textsuperscript{3} and A. Whiteway\textsuperscript{2}}
\IEEEauthorblockA{\small \textsuperscript{1}Centre for Security, Theory and Algorithmic Research, International Institute of Information Technology, Hyderabad, India\\
\textsuperscript{2}Risk Analytics Laboratory, Faculty of Computer Science, Dalhousie University, Halifax, Canada\\
\textsuperscript{3}Big Data Laboratory, School of Computer Science, University of St Andrews, Scotland, UK}\\
aman.kumar@research.iiit.ac.in, {obaltzer, arc}@cs.dal.ca, varghese@st-andrews.ac.uk, aaron.whiteway@dal.ca
}
\maketitle

\begin{abstract}
Stochastic simulation techniques employed for the analysis of portfolios of insurance/reinsurance risk, often referred to as `Aggregate Risk Analysis', can benefit from exploiting state-of-the-art high-performance computing platforms. In this paper, parallel methods to speed-up aggregate risk analysis for supporting real-time pricing are explored. An algorithm for analysing aggregate risk is proposed and implemented for multi-core CPUs and for many-core GPUs. Experimental studies indicate that GPUs offer a feasible alternative solution over traditional high-performance computing systems. A simulation of 1,000,000 trials with 1,000 catastrophic events per trial on a typical exposure set and contract structure is performed in less than 5 seconds on a multiple GPU platform. The key result is that the multiple GPU implementation can be used in real-time pricing scenarios as it is approximately \textbf{77x} times faster than the sequential counterpart implemented on a CPU.
\end{abstract}

\begin{IEEEkeywords}
GPU computing; aggregate risk analysis; catastrophe event risk; real-time pricing

\end{IEEEkeywords}

\IEEEpeerreviewmaketitle

\section{Introduction}
\label{introduction}
Large-scale simulations in the risk analytics domain \cite{riskanalytics1, riskanalytics2} are both data and computationally intensive. They can benefit from exploiting advances in high-performance computing. While a large number of financial engineering applications, for example \cite{financialengineering1, financialengineering2} are benefiting from the advancement of high-performance computing, there are relatively fewer insurance and reinsurance applications exploiting parallelism. In this paper, we explore parallel methods and their implementations for aggregate risk analysis \cite{agganalysis1, agganalysis4} required in portfolio risk management and real-time pricing of insurance and reinsurance contracts.

Aggregate risk analysis is a form of Monte Carlo simulation performed on a portfolio of risks that an insurer or a re-insurer holds rather than on individual risks. A portfolio may comprise tens of thousands of contracts that cover risks associated with catastrophic events such as earthquakes, hurricanes and floods. Generally, contracts have an `eXcess of Loss' (XL) \cite{excessloss1} structure and may provide coverage for single event occurrences up to a specified limit with an optional retention by the insured, or for multiple event occurrences up to a specified aggregate limit with an optional retention by the insured, or for a combination of both. Each trial in the aggregate risk analysis simulation represents a view of the occurrence of catastrophic events and the order in which they occur within a predetermined period, (i.e., a contractual year) and how they will interact with complex treaty terms to produce an aggregated loss. 

From a computational perspective the aggregate risk analysis simulation differs from other Monte Carlo simulations since trials are pre-simulated, rather than randomly generated on-the-fly. This provides millions of alternate views of a contractual year comprising thousands of events which are pre-simulated as a Year Event Table (YET). From an analytical perspective a pre-simulated YET lends itself to statistical validation and to tuning for seasonality and cluster effects. Although such a simulation provides actuaries and decision makers with a consistent lens to view results, there are significant challenges in achieving efficient parallelisation. The extremely large YET must be carefully shared between processing cores if the computation is to achieve reasonable speed-up in the face of limited memory bandwidth.

With inputs, namely the YET, a portfolio of contracts and a set of Event Loss Tables, the output of aggregate analysis is a Year Loss Table (YLT). From a YLT, an insurer or a re-insurer can derive important portfolio risk metrics, such as the Probable Maximum Loss (PML) \cite{pml1, pml2} and the Tail Value-at-Risk (TVaR) \cite{tvar1, tvar2}, which are used for internal risk management and reporting to regulators and rating agencies. 

In this paper, firstly, a sequential aggregate risk analysis algorithm is proposed and implemented in C++ on a CPU, followed by parallel implementations using C++ and OpenMP on a multi-core CPU and using C++ and CUDA on many-core GPU platforms. The algorithms must ingest large amounts of data in the form of the Year Event Table and the Event Loss Tables, and therefore, challenges such as efficiently organising input data in limited memory, and defining the granularity at which parallelism can be applied to the problem for achieving speed-up are considered. 

Preliminary efforts \cite{sc12} on single GPU methods for this problem achieve some speed-up but did not evaluate some of the more important GPU optimisation methods \cite{gpu1, gpu2}. Optimisations, such as chunking, loop unrolling, reducing the precision of variables used and the usage of kernel registries over shared and global memories of the GPU are performed in this paper to improve the speed-up achieved on the GPU; the result is a maximum speed-up of 77x which is achieved for the parallel implementation on the multiple GPU consisting of a CPU and four GPUs as compared to the sequential implementation on a CPU. The results indicate the feasibility of employing aggregate risk analysis on multiple GPUs for real-time pricing scenarios. The implementations presented in this paper are cost effective high-performance computing solutions over massive conventional clusters and supercomputers. The GPU implementations takes full advantage of the high levels of parallelism, some advantage of fast shared memory access, but surprisingly little advantage of the fast numerical performance all offered by the machine architecture of GPUs. 

The remainder of this paper is organised as follows. Section \ref{aggregateriskanalysis} presents the aggregate risk analysis algorithm, its inputs and outputs. Section \ref{experimentalstudies} considers the implementations of the algorithm on a multi-core CPU and a many-core GPU. Section \ref{performanceanalysis} highlights the results obtained from an analysis of the performance of the algorithm. Section \ref{discussion} compares and contrasts the algorithms and the results obtained from the experiments. Section \ref{conclusion} concludes the paper by considering future work. 

\section{Aggregate Risk Analysis}
\label{aggregateriskanalysis}
The inputs and the algorithm for aggregate risk analysis are considered in this section. There are three inputs to the procedure that analyses aggregate risk. The first input is a database of pre-simulated occurrences of events from a catalogue of stochastic events, which is referred to as the Year Event Table ($YET$). A possible sequence of catastrophe event occurrences for any given year is defined as a record in the $YET$ is a `trial' ($T_{i}$). The sequence of events is defined by a set of tuples containing the ID of an event and the time-stamp of its occurrence in a trial $T_i = \{(E_{i, 1}, t_{i, 1}), \dots, (E_{i, k}, t_{i, k})\}$ 
which is ordered by ascending time-stamp values. A typical YET may comprise thousands to millions of trials, and each trial may have approximately between 800 to 1500 `event time-stamp' pairs, based on a global event catalogue covering multiple perils. The $YET$ is represented as 
\begin{center}
$YET=\left\{ T_i = \left\{(E_{i, 1}, t_{i, 1}), \dots, (E_{i, k}, t_{i, k})\right\} \right\}$,\\ %\vspace{6pt}
where $i = 1, 2, \dots$ and $k = 1, 2, \dots, 800-1500$
\end{center}

The second input is a collection of specific events and their corresponding losses with respect to an exposure set referred to as the Event Loss Tables ($ELT$). An event may be part of multiple ELTs and associated with a different loss in each ELT. For example, one ELT may contain losses derived from one exposure set while another ELT may contain the same events but different losses derived from a different exposure set. Each ELT is characterised by its own meta data including information about currency exchange rates and terms that are applied at the level of each individual event loss. Each record in an ELT is denoted as event loss $EL_{i} = \{E_{i}, l_{i}\}$, and the financial terms associated with the ELT are represented as a tuple $\mathcal{I} = (\mathcal{I}_{1}, \mathcal{I}_{2}, \dots)$. A typical aggregate analysis may involve 10,000 ELTs, each containing 10,000-30,000 event losses with exceptions even up to 2,000,000 event losses. The ELTs are represented as 
\begin{center}
$ELT=\left\{\begin{array}{c}EL_{i} = \{E_{i}, l_{i}\}, \\ \mathcal{I} = (\mathcal{I}_{1}, \mathcal{I}_{2}, \dots)\end{array}\right\}$,\\ \vspace{6pt}
with $i = 1, 2, \dots , 10,000-30,000$
\end{center}

The third input is Layers, denoted as $L$, which cover a collection of ELTs under a set of layer terms. Each layer $L_{i}$ defines a single reinsurance contract and consists two attributes. Firstly, the set of ELTs $\mathcal{E} = \{ELT_1, ELT_2, \dots, ELT_j\}$ to be covered by the layer, and secondly, the Layer Terms, denoted as $\mathcal{T} = (\mathcal{T}_{OccR}, \mathcal{T}_{OccL}, \mathcal{T}_{AggR}, \mathcal{T}_{AggL})$ which defines the contractual terms. A typical layer covers approximately 3 to 30 individual ELTs. A Layer is represented as 
\begin{center}
$L=\left\{\begin{array}{c}\mathcal{E}=\{ELT_1,ELT_2,\dots,ELT_j\},\\ \mathcal{T}=(\mathcal{T}_{OccR},\mathcal{T}_{OccL},\mathcal{T}_{AggR},\mathcal{T}_{AggL})\end{array}\right\}$,\\ \vspace{6pt}
with $j = 1, 2,\dots, 3-30$.
\end{center}

The algorithm (line no. 1-32 shown in Algorithm \ref{algorithm1}) for aggregate analysis consists of two stages. In the first stage, referred to as the preprocessing stage, data (the $YET$, $ELT$ and $L$) is loaded into local memory. 

\begin{algorithm}
\caption{Aggregate Risk Analysis}
\label{algorithm1}
\begin{algorithmic}[1]

%\Procedure{AggregateRiskAnalysis}{$YET$, $ELT$, $L$}
\Procedure{ARA}{$YET$, $ELT$, $L$}

\ForAll{$a \in L$} 
	\ForAll{$b \in YET$}
		\ForAll{$c \in (EL \in a)$}
			\ForAll{$d \in (Et \in b)$}
				\State $x_{d} \Leftarrow E \in d$ in $El \in f$, where $f \in ELT$ and $(EL \in f) = c$
			\EndFor
			\ForAll{$d \in (Et \in b)$}
				\State $l_{x_{d}} \Leftarrow Apply Financial Terms(\mathcal{I})$
			\EndFor
			\ForAll{$d \in (Et \in b)$}
				\State $lo_{x_{d}} \Leftarrow lo_{x_{d}} + l_{x_{d}}$
			\EndFor				
		\EndFor	
		\ForAll{$d \in (Et \in b)$}
			\State $lo_{x_{d}} \Leftarrow min(max(lo_{x_{d}} - \mathcal{T}_{OccR}, 0), \mathcal{T}_{OccL})$
		\EndFor
		\ForAll{$d \in (Et \in b)$}
			\State $lo_{x_{d}} \Leftarrow \sum\limits_{i=1}^{d}lo_{x_{i}}$
		\EndFor
		\ForAll{$d \in (Et \in b)$}
			\State $lo_{x_{d}} \Leftarrow min(max(lo_{x_{d}} - \mathcal{T}_{AggR}, 0), \mathcal{T}_{AggL})$
		\EndFor
		\ForAll{$d \in (Et \in b)$}
			\State $lo_{x_{d}} \Leftarrow lo_{x_{d}} - lo_{x_{d - 1}}$
		\EndFor
		\ForAll{$d \in (Et \in b)$}
			\State $lr \Leftarrow lr + lo_{x_{d}}$
		\EndFor
	\EndFor
\EndFor
\EndProcedure
\end{algorithmic}
\end{algorithm}

In the second stage, a four step simulation for each Layer and for each trial in the YET is performed and a Year Loss Table ($YLT$) is produced. Line no. 4-7 shows the first step in which each event of a trial its corresponding event loss in the set of ELTs associated with the Layer is determined.

Line no. 8-10 shows the second step in which a set of financial terms is applied to each event loss pair extracted from an ELT. In other words, contractual financial terms to the benefit of the layer are applied in this step. For this the losses for a specific event's net of financial terms $\mathcal{I}$ are accumulated across all ELTs into a single event loss shown in line no. 11-13.

Line no. 15-20 shows the third step in which the event loss for each event occurrence in the trial, combined across all ELTs associated with the layer, is subject to occurrence terms (i) Occurrence Retention, denoted as $\mathcal{T}_{OccR}$, which is the retention or deductible of the insured for an individual occurrence loss, and (ii) Occurrence Limit, denoted as $\mathcal{T}_{OccL}$, which is the limit or coverage the insurer will pay for occurrence losses in excess of the retention. Occurrence terms are applicable to individual event occurrences independent of any other occurrences in the trial. The occurrence terms capture specific contractual properties of 'eXcess of Loss' treaties as they apply to individual event occurrences only. The event losses net of occurrence terms are then accumulated into a single aggregate loss for the given trial.

Line no. 21-29 shows the fourth step in which the aggregate terms (i) Aggregate Retention, denoted as $\mathcal{T}_{AggR}$, which is the retention or deductible of the insured for an annual cumulative loss, and (ii) Aggregate Limit, denoted as $\mathcal{T}_{AggL}$, which is the limit or coverage the insurer will pay for annual cumulative losses in excess of the aggregate retention. Aggregate terms are applied to the trial's aggregate loss for a layer. Unlike occurrence terms, aggregate terms are applied to the cumulative sum of occurrence losses within a trial and thus the result depends on the sequence of prior events in the trial. This behaviour captures contractual properties as they apply to multiple event occurrences. The aggregate loss net of the aggregate terms is referred to as the trial loss or the year loss and stored in a Year Loss Table (YLT) as the result of the aggregate analysis.

The algorithm will provide an aggregate loss value for each trial denoted as $lr$ in line no. 28. Financial functions or filters are then applied on the aggregate loss values.

\section{Experimental Studies}
\label{experimentalstudies}
The experimental studies investigate the sequential and parallel implementation of the aggregate risk analysis on three hardware platforms. Firstly, a multi-core CPU is employed whose specifications are a 3.40 GHz quad-core Intel(R) Core (TM) i7-2600 processor with 16.0 GB of RAM. The processor had 256 KB L2 cache per core, 8 MB L3 cache and maximum memory bandwidth of 21 GB/sec. Both sequential and parallel versions of the aggregate risk analysis algorithm were implemented on this platform. The sequential version was implemented in C++, while the parallel version was implemented in C++ and OpenMP. Both versions were compiled using the GNU Compiler Collection g++ 4.4 using the `\texttt{-O3}' flag.

Secondly, a NVIDIA Tesla C2075 GPU, consisting of 448 processor cores (organised as 14 streaming multi-processors each with 32 symmetric multi-processors), each with a frequency of 1.15 GHz, a global memory of 5.375 GB and a memory bandwidth of 144 GB/sec was employed in the GPU implementations of the aggregate risk analysis algorithm. The peak double precision floating point performance is 515 Gflops whereas the peak single precision floating point performance is 1.03 Tflops. 

Thirdly, a multiple GPU platform comprising four NVIDIA Tesla M2090 GPUs, and each GPU consists 512 processor cores (organised as 14 streaming multi-processors each with 32 symmetric multi-processors) and 5.375 GB of global memory with a memory bandwidth of 177 GB/sec is employed for implementing the fastest aggregate risk analysis algorithm reported in this paper. The peak double precision floating point performance is 665 Gflops whereas the peak single precision floating point performance is 1.33 Tflops. CUDA is employed for the basic GPU implementation of the aggregate risk analysis algorithm and the optimised implementations.

Five variations of the algorithm are implemented, they are: (i) a sequential implementation, (ii) a parallel implementation for multi-cores CPUs, (iii) a parallel GPU implementation, (iv) an optimised parallel implementation on the GPU, and (v) an optimised parallel implementation on a multiple GPU.

In all implementations a single thread is employed per trial, $T_{id}$. The key design decision from a performance perspective is the selection of a data structure for representing Event Loss Tables (ELTs). ELTs are essentially dictionaries consisting of key-value pairs and the fundamental requirement is to support fast random key lookup. The ELTs corresponding to a layer were implemented as direct access tables. A direct access table is a highly sparse representation of a ELT, one that provides very fast lookup performance at the cost of high memory usage. For example, consider an event catalogue of 2,000,000 events and a ELT consisting of 20,000 events for which non-zero loss values were obtained. To represent the ELT using a direct access table, an array of 2,000,000 loss values are generated in memory of which 20,000 are non-zero loss values and the remaining 1,980,000 events have zero loss values. So if a layer has 15 ELTs, then $15 \times 2,000,000 = 30,000,000$ event-loss pairs are generated in memory.

A direct access table was employed in all implementations over any alternate compact representation for the following reasons. A search operation is required to find an event-loss pair in a compact representation. Sequential and binary search require $O(n)$ and $O(log(n))$ memory accesses respectively to locate an event-loss pair. Even if a constant-time space-efficient hashing scheme (for example, cuckoo hashing \cite{cuckoohashing}) requiring a constant number of memory accesses is adopted there is considerable implementation and run-time performance complexity. This overhead will be high on GPUs with complex memory hierarchies consisting of global and shared memories. To perform aggregate analysis on a YET of 1 million trials (each trial comprising 1000 events) and for a layer covering 15 ELTs, there are $1,000 \times 1,000,000 \times 15 = 15,000,000,000$ events, which require random access to 15 billion loss values. Direct access tables, although wasteful of memory space, allow for the fewest memory accesses as each lookup in an ELT requires only one memory access per search operation.

Two data structure implementations of the 15 ELTs were considered. In the first implementation, each ELT is an independent table, and therefore, in a read cycle, each thread independently looks up its events from the ELTs. All threads within a block access the same ELT. By contrast, in the second implementation, the 15 ELTs are combined as a single table. Consequently, the threads then use the shared memory to load entire rows of the combined ELTs at a time. The second implementation has comparatively poorer performance than the first because for the threads to collectively load from the combined ELT each thread must first write which event it needs. This results in additional memory overheads. 

In the basic implementation on the multi-core CPU platform the entire data required for the algorithm is processed in memory. The GPU implementation of the basic algorithm uses the GPU's global memory to store all of the required data structures. The basic parallel implementation on the GPU requires high memory transactions and leads to inefficient performance on the GPU platform. To surmount this challenge shared memory can be utilised over global memory. 

The implementation on the GPU builds on the parallel implementation and considers three optimisations. Firstly, chunking, which refers to processing a block of events of fixed size (or chunk size) for the efficient use of shared memory. The four steps (lines 4-29 in the basic algorithm, i.e., events in a trial and both financial and layer terms computations) of the algorithm are chunked. In addition, the financial and layer terms are stored in the streaming multi-processor's constant memory. In the basic implementation, $l_{x_{d}}$ and $lo_{x_{d}}$ are represented in the global memory and therefore, in each step while applying the financial and layer terms the global memory has to be accessed and updated adding considerable overhead. This overhead is minimised in the optimised implementation by (a) chunking the financial and layer term computations, and (b) chunking the memory read operations (line no. 4-7) for reading events in a trial from the $YET$, represented by $Et_{id}$. Chunking reduces the number of global memory update and global read operations. Moreover, the benefits of data striding can also be used to improve speed-up. 

Secondly, loop unrolling, which refers to the replication of blocks of code included within for loops by the compiler to reduce the number of iterations performed by the for loop. The for loops provided in lines 5 and 8 are unrolled using the pragma directive, thereby reducing the number of instructions that need to be executed by the GPU. 

%Thirdly, reducing the precision of variables, whereby the double variables are changed to float variables. Read operations are faster using float variables as they are only half the size of a double variable. Furthermore, the performance of single precision operations tend to be approximately twice as fast as double precision operations on GPUs.

Thirdly, migrating data from both shared and global memory to the kernel registry. The kernel registry has the lowest latency compared to all other memory.

The optimised aggregate risk analysis algorithm was also implemented on a multiple GPU platform. This implementation was achieved by decomposing the aggregate analysis workload among the four available GPUs. For this a thread on the CPU invokes and manages a GPU. The CPU thread calls a method which takes as input all the inputs required by the kernel (the three inputs are presented in Section \ref{aggregateriskanalysis}) and the pre-allocated arrays for storing the outputs generated by the kernel. The CPU threads are invoked in a parallel manner thereby contributing to the speed-up achieved on the multiple GPU platform. 

\section{Performance Analysis}
\label{performanceanalysis}
In this section, the results obtained from the sequential and parallel implementations are considered. 

\subsection{Results from multi-core CPU}
In the experiments for the sequential implementation it was observed that there is a linear increase on running time of executing the sequential version of the basic aggregate analysis algorithm on a CPU using a single core when the number of the number of events in a trial, number of trials, average number of ELTs per layer and number of layers is increased. For a typical aggregate analysis problem comprising 1,000,000 trials and each trial comprising 1,000 events the sequential algorithm takes 337.47 seconds, with over 65\% of the time for look-up of Loss Sets in the direct access table, and with only over 31\% of the time for the numerical computations. This indicates that in addition to improving the speed of the numerical computations, techniques to lower the time for look-up can provide significant speedup in the parallel implementations. 

\begin{figure} [t] %[!tp]
\centering
	\subfloat[No. of cores vs execution time]{\label{fig:s21}\includegraphics[width=0.485\textwidth]{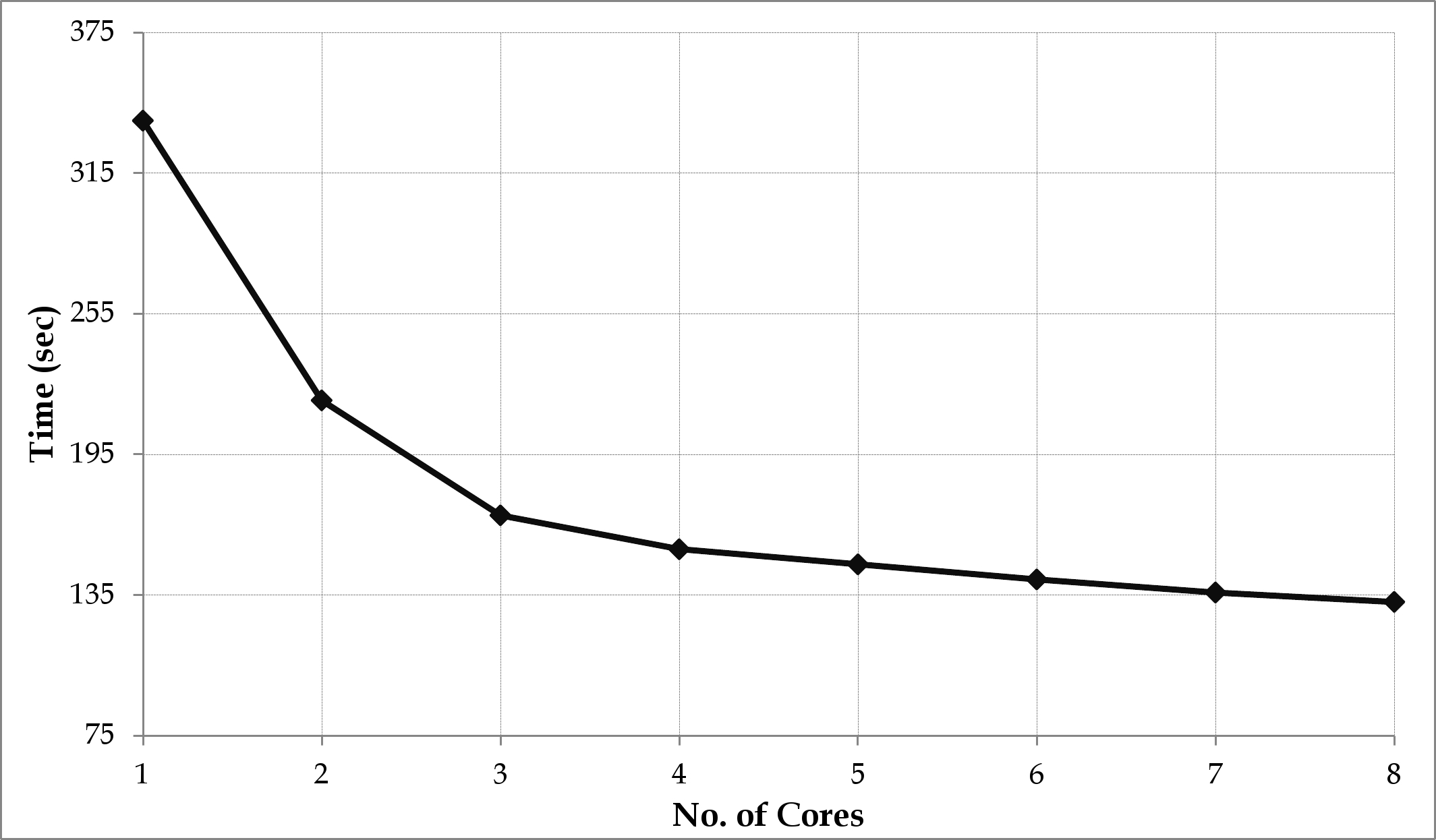}} \\%\hfill
	\subfloat[Total No. of threads vs execution time]{\label{fig:s22}\includegraphics[width=0.485\textwidth]{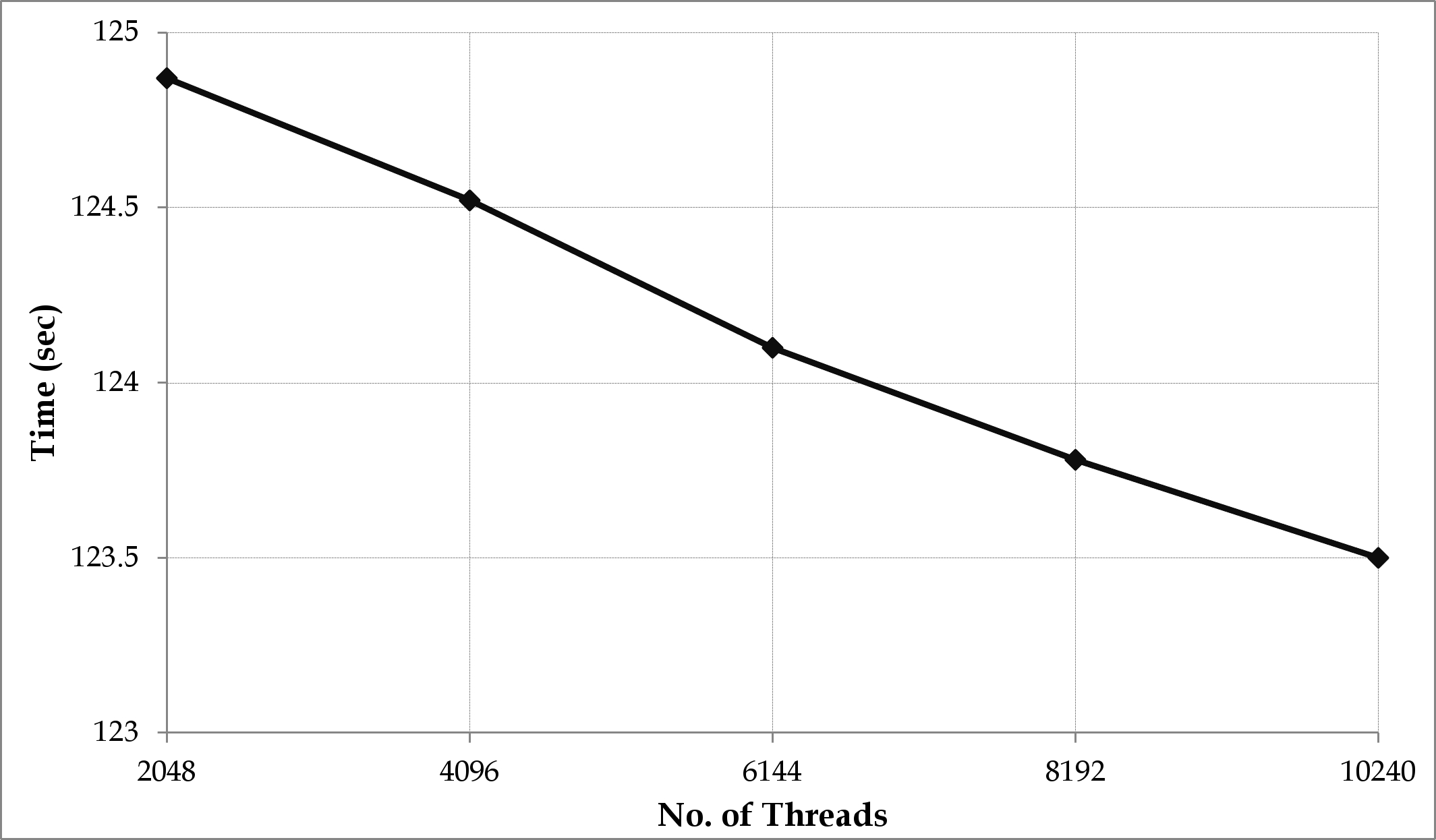}} \\
\caption{Performance of the parallel implementation of the aggregate analysis algorithm on a multi-core CPU}
\label{graphset1}
\end{figure}

Figure \ref{graphset1} illustrates the performance of the basic aggregate analysis algorithm on a multi-core CPU. In Figure \ref{fig:s21}, a single thread is run on each core and the number of cores is varied from 1 to 8. Each thread performs aggregate analysis for a single trial and threading is implemented by introducing OpenMP directives into the C++ source. Limited speed-up is observed. For two cores we achieve a speed-up of 1.5x, for four cores the speed-up is 2.2x, and for 8 cores it is only 2.6x. As we increase the number of cores we do not equally increase the bandwidth to memory which is the limiting factor. The algorithm spends most of its time performing random access reads into the ELT data structures. Since these accesses exhibit no locality of reference they are not aided by the processors cache hierarchy. A number of approaches were attempted, including the chunking method for GPUs, but were not successful in achieving a high speed-up on our multi-core CPU. However a moderate reduction in absolute time by running many threads on each core was achieved.

Figure \ref{fig:s22} illustrates the performance of the basic aggregate analysis engine when all 8 cores are used and each core is allocated many threads. As the number of threads are increased an improvement in the performance is noted. With 256 threads per core (i.e. 2048 in total) the overall runtime drops from 135 seconds to 125 seconds. Beyond this point we observe diminishing returns as illustrated in Figure \ref{fig:s21}.

\subsection{Results from many-core GPU}
In the GPU implementations, CUDA provides an abstraction over the streaming multi-processors, referred to as a CUDA block. When implementing the basic aggregate analysis algorithm on a GPU we need to select the number of threads executed per CUDA block.  For example, consider 1 million threads are used to represent the simulation of 1 million trials on the GPU, and 256 threads are executed on a streaming multi-processor. There will be $\frac{1,000,000}{256} \approx 3906$ blocks in total which will have to be executed on 14 streaming multi-processors. Each streaming multi-processor will therefore have to execute $\frac{3906}{14} \approx 279$ blocks. Since the threads on the same streaming multi-processor share fixed size allocations of shared and constant memory there is a real trade-off to be made. If we have a smaller number of threads, each thread can have a larger amount of shared and constant memory, but with a small number of threads we have less opportunity to hide the latency of accessing the global memory.

\begin{figure} %[t]
\centering
	\includegraphics[width=0.485\textwidth]{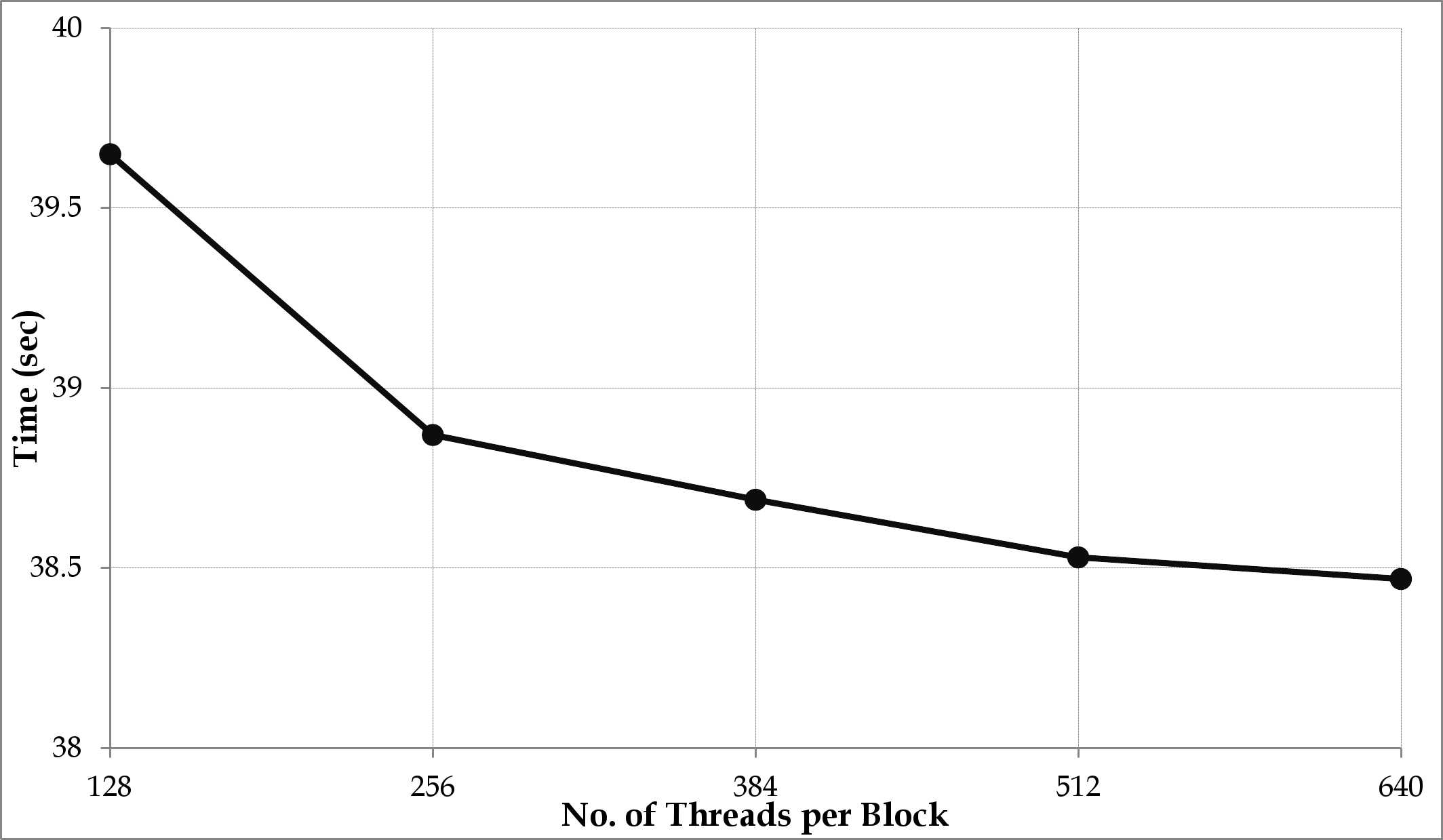}
	\caption{Graphs plotted for number of threads vs the time taken for executing the parallel implementation on many-core GPU}
	\label{graph2}
\end{figure}

Figure \ref{graph2} shows the time taken for executing the parallel version of the basic implementation on the GPU when the number of threads per CUDA block are varied between 128 and 640. At least 128 threads per block are required to efficiently use the available hardware. An improved performance is observed with 256 threads per block but beyond that point the performance improvements diminish greatly.

The optimised implementation of the aggregate risk analysis algorithm on the GPU platform aims to utilise shared and constant memory as much as possible by processing ``chunks'', blocks of events of fixed size (referred to as chunk size), to improve the utilisation of the faster shared memories that exist on each streaming multi-processor. Further to optimise the implementation, loops are unrolled, the precision of variables are reduced by changing the double variables to float variables, and data from both shared and global memory are migrated to the kernel registry. The optimised algorithm has a significantly reduced runtime from 38.47 seconds down to 20.63 seconds, representing approximately a 1.9x improvement. 

\subsection{Results from multiple GPUs}
\begin{figure} [t] %[!tp]
\centering
	\subfloat[No. of GPUs vs execution time]{\label{fig:s41}\includegraphics[width=0.485\textwidth]{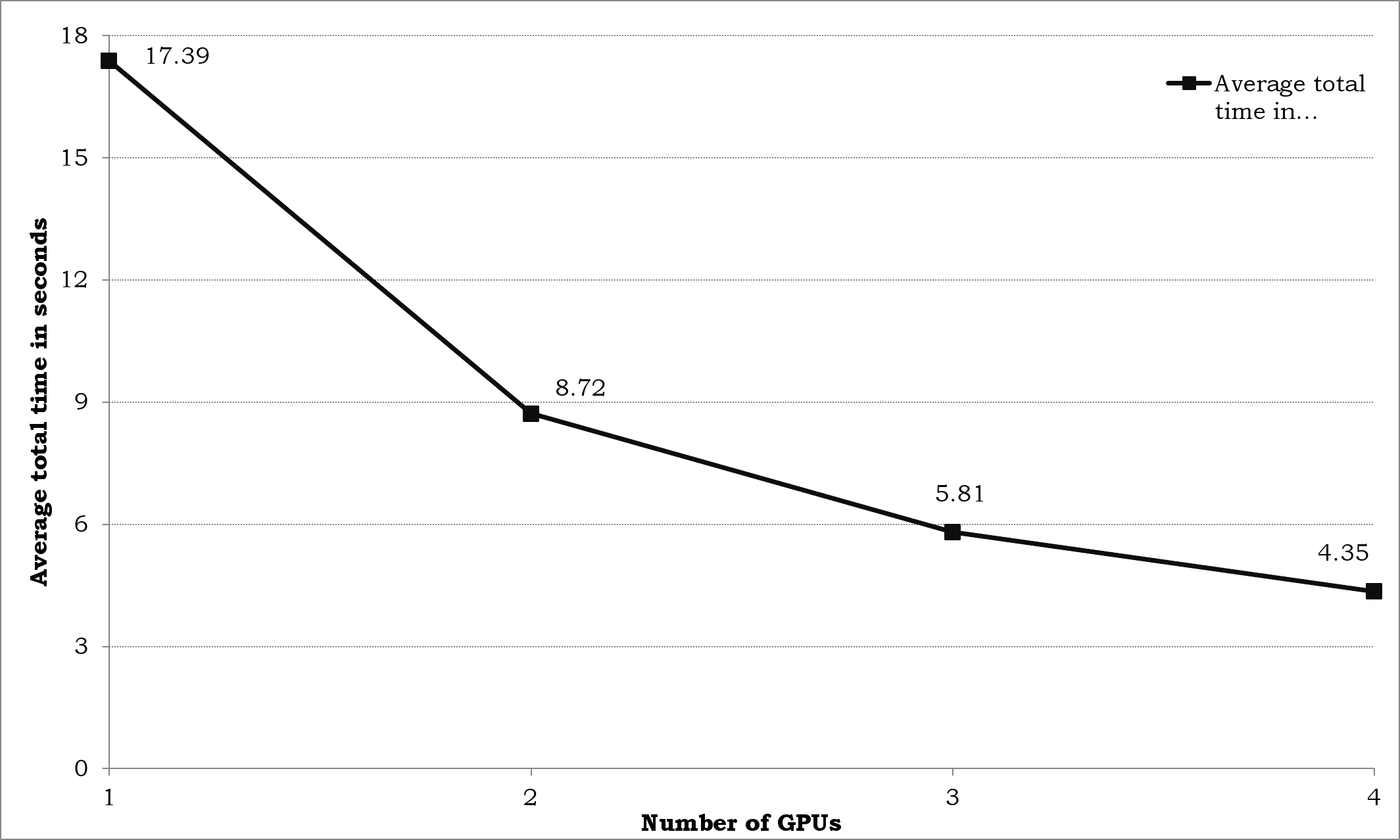}} \\%\hfill
	\subfloat[Efficiency of the implementation on multiple GPUs]{\label{fig:s42}\includegraphics[width=0.485\textwidth]{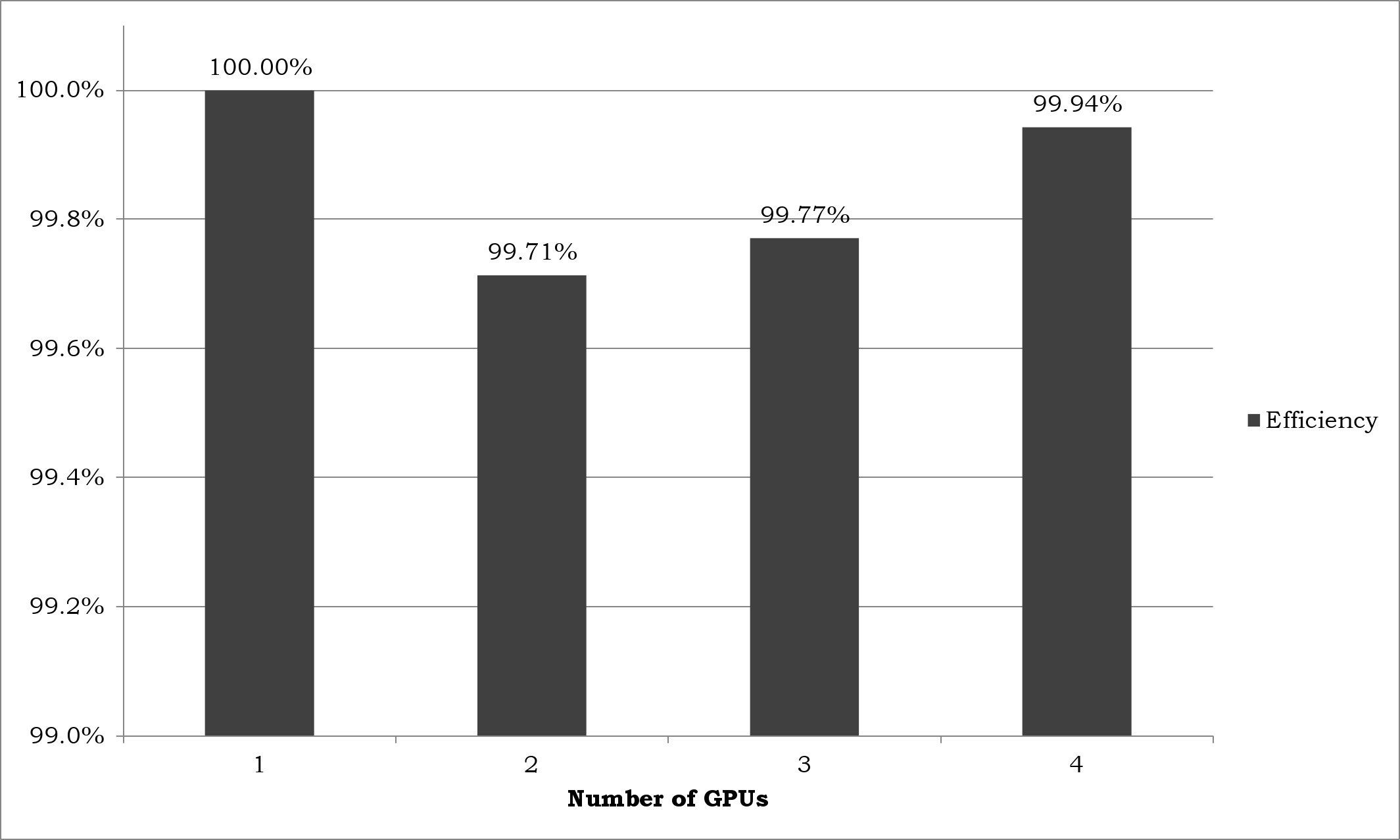}} \\
\caption{Performance of the optimised parallel implementation of the aggregate analysis algorithm on multiple GPUs}
\label{graphset4}
\end{figure}

Figure \ref{graphset4} illustrates the performance of the optimised aggregate analysis algorithm on multiple GPUs. A CPU thread is used to select an available GPU for executing the aggregate analysis problem that is decomposed. In the experiments, aggregate analysis algorithm is executed using one, two, three and four GPUs. A much higher speed-up is achieved on the multiple GPU over single GPU; the time taken for look-up of Loss Sets in the direct access table drops from 20.1 seconds to 4.25 seconds and the time for all Financial Term and Layer Term computations drop from 0.11 seconds to 0.02 seconds. The best average time obtained for executing the optimised algorithm on four GPUs is 4.35 seconds which is around 5x times faster than the time taken on the many-core GPU and 4x times faster than the time taken by the implementation executing on a single GPU of the multiple GPU machine (refer Figure \ref{fig:s41}). The results from the multiple GPU show approximately 100\% efficiency (refer Figure \ref{fig:s42}). 

\begin{figure} %[t]
\centering
	\includegraphics[width=0.485\textwidth]{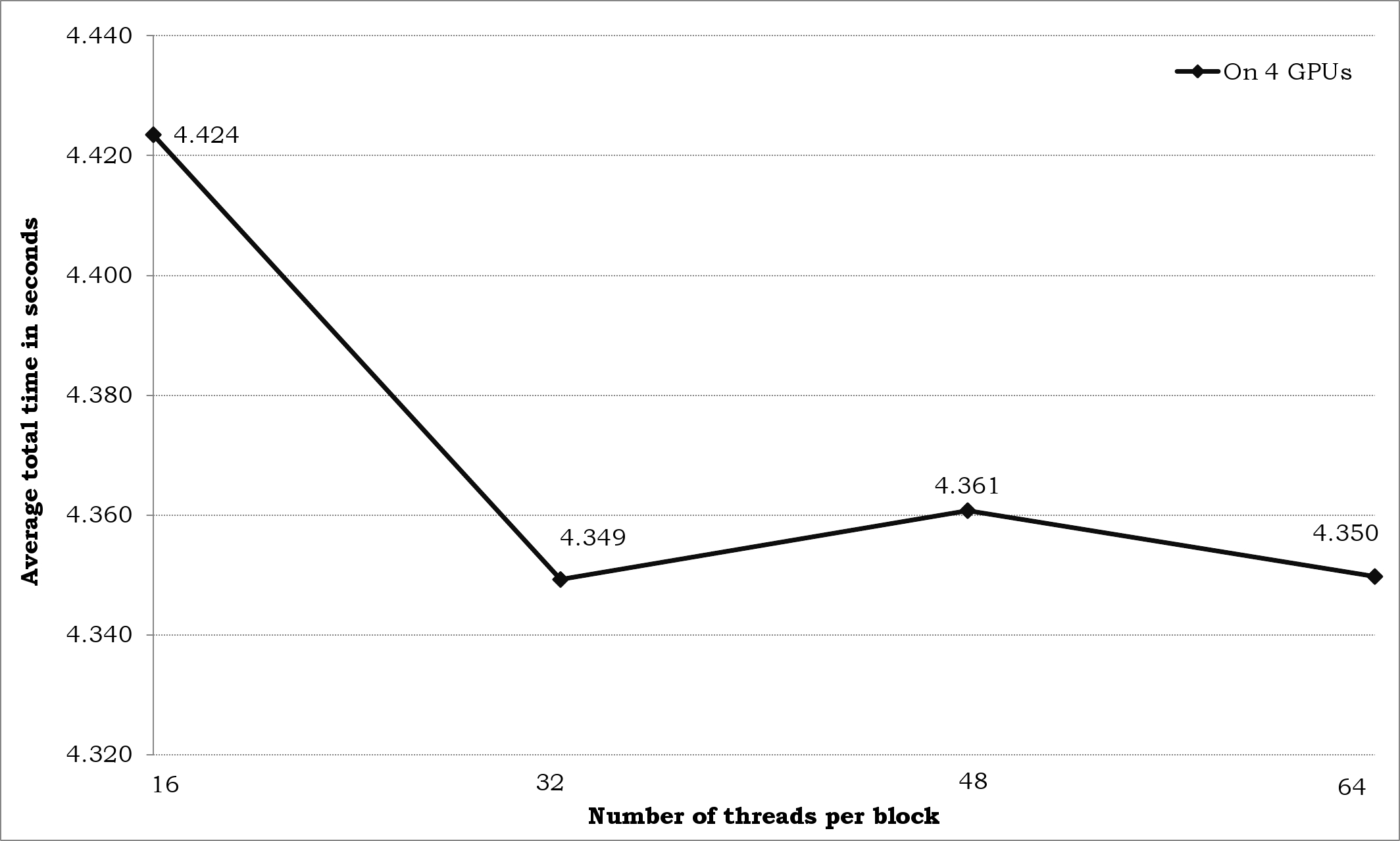}
	\caption{Graphs plotted for the number of threads per block size for four GPUs vs the time taken for executing the optimised parallel implementation on multiple GPUs}
	\label{graph5}
\end{figure}

Figure \ref{graph5} shows the performance of the optimised aggregate analysis algorithm on four GPUs when the number of threads per block is varied from 16 to 64. Experiments could not be pursued beyond 64 threads per block due to the limitation on the block size the shared memory can use. The best performance of 4.35 seconds is achieved when the number of threads per block is 32; this is so as the block size is the same as the WARP size of the GPU whereby an entire block of threads can be swapped when high latency operations occur. Increasing the number of threads per block does not improve the performance owing to shared memory overflow. 

\section{Discussion}
\label{discussion}

\begin{figure} [t]
\centering
	\includegraphics[width=0.485\textwidth]{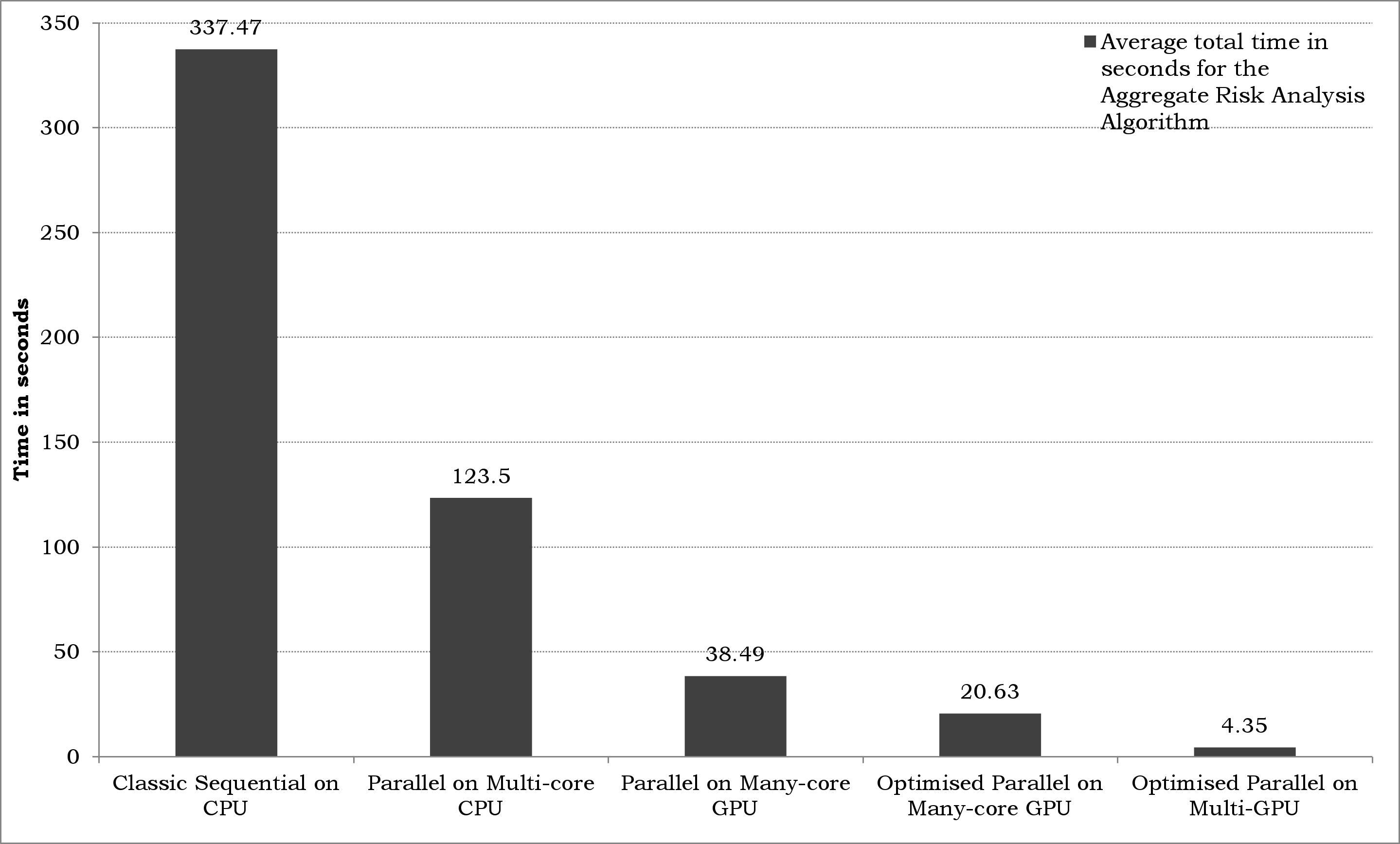}
	\caption{Bar graphs plotted for the average total time for executing the sequential and parallel implementations on the hardware computing platforms}
	\label{graph10}
\end{figure}

\begin{figure*} [t]
\centering
	\includegraphics[width=0.995\textwidth]{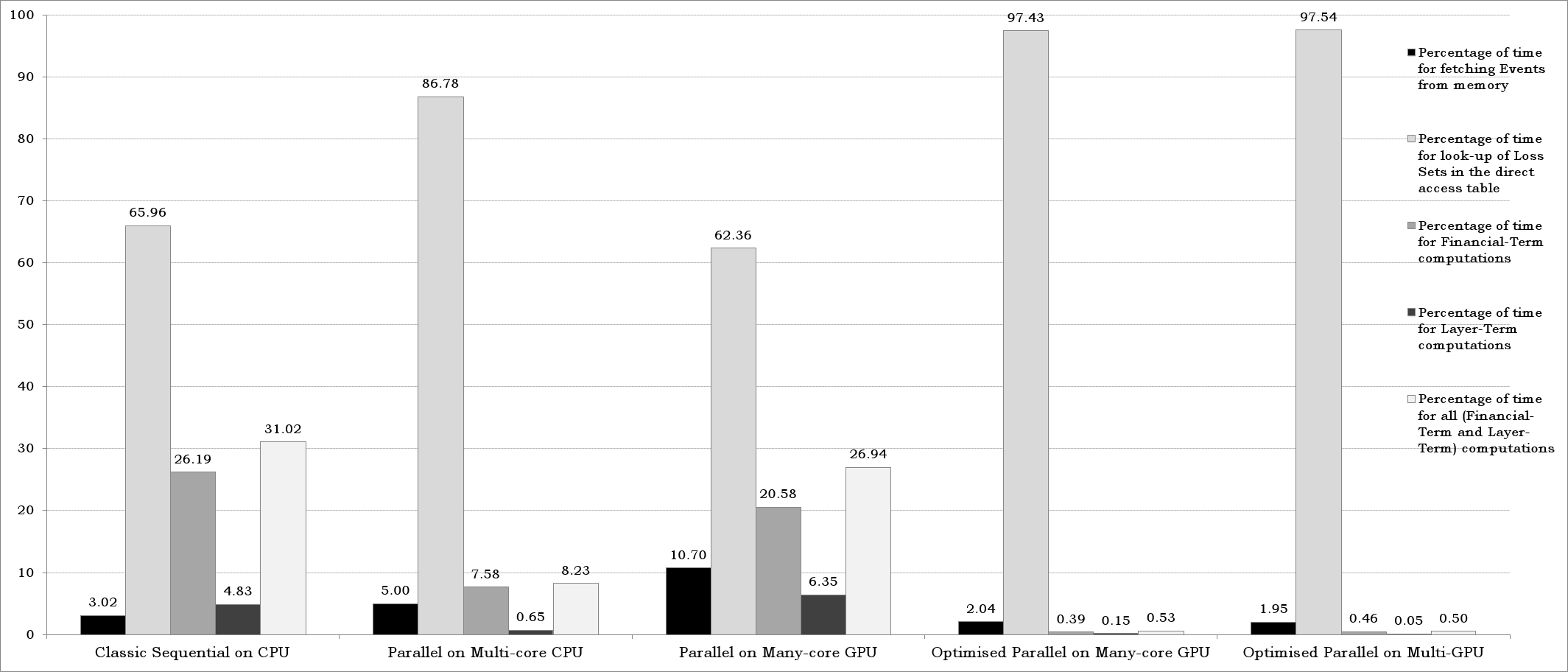}
	\caption{Bar graphs plotted for the percentages of time for different activities in aggregate risk analysis}
	\label{graph11}
\end{figure*}

Figure \ref{graph10} and Figure \ref{graph11} summarises the results obtained from all the experiments of (a) a sequential implementation on the CPU, (b) a parallel implementation on the multi-core CPU, (c) a parallel implementation on the many-core GPU, (d) an optimised parallel implementation on the many-core GPU, and (e) an optimised parallel implementation on the multiple GPU. 

Figure \ref{graph10} shows the decrease in the total time taken for executing the aggregate analysis problem for 1 Layer, 15 Loss Sets and 1,000,000 Trials with 1,000 catastrophic Events per Trial from 337.47 seconds for a sequential implementation on the CPU to just 4.35 seconds for an optimised parallel implementation on four GPUs. The parallel implementation on the multi-core CPU takes 123.5 seconds which is approximately one-third the time taken for the sequential implementation. This speed-up is due to the use of multiple cores of the CPU, and there are memory limitations to achieve any further speed-up. The time taken for executing the parallel implementation on the many-core GPU is reduced further by approximately one-third to 38.49 seconds over the multi-core CPU. This speed-up is achieved due to the GPU architecture which offers lots of cycles for independent parallelism, fast memory access and fast numerical computations. The time taken for executing the optimised parallel implementation on the many-core GPU is reduced further by approximately half to 20.63 seconds over the many-core GPU. The speed-up achieved in this case is attributed to four optimisations in the form of (i) chunking, (ii) loop unrolling, (iii) reducing the precision of variables and (iv) migrating data to kernel registry. The optimised parallel implementation on the multiple GPU takes 4.35 seconds which is approximately 77 times faster than the CPU; the speed-up in this case is achieved due to optimisations and the use of multiple GPUs. 

Figure \ref{graph11} shows the percentage of time taken for (a) fetching Events from memory, (b) look-up of Loss Sets in direct access table, (c) Financial-Term computations, (d) Layer-Term computations, and (e) both Financial-Term and Layer-Term computations. The best time taken for fetching Events from memory in the sequential implementation on the CPU is over 10 seconds, in the parallel implementation on the multi-core CPU is nearly 6 seconds, in the parallel implementation on the many-core GPU is nearly 4 seconds, in the optimised implementation on the many-core GPU is less than 0.5 seconds and on the multiple GPU is less than 0.1 seconds. Precisely the most optimised implementation on the multiple GPU has an improvement of 100 times for the time taken in fetching Events from memory over the sequential implementation on the CPU. 

The majority of the total time taken for executing the aggregate analysis problem is for the look-up of Loss Sets in the direct access table. While the sequential implementation requires 222.61 seconds for the look-up, the optimised implementation on the multiple GPU only requires 4.25 seconds, which is an improvement of 50 times. However, there is scope for improvement to bring down the time. Surprisingly, on the multiple GPU 97.54\% of the total time (4.33 seconds) is for look-up. This calls for exploring optimised techniques for look-up in the direct access table to further reduce the overall time taken for executing the aggregate analysis problem. 

The numerical computations, including both the Financial-Term and Layer-Term computations take 104.67 seconds for the sequential implementation on the CPU and only one-tenth that time for the parallel implementation on the CPU. The most optimised implementation takes merely 0.02 seconds on the multiple GPU platform which is approximately 5000 times faster than the sequential implementation on the GPU. The cutting edge technology offered by GPU architectures for numerical computations is fully harnessed to significantly lower the computational time in the aggregate analysis algorithm. 

To summarise, the results obtained by profiling of time indicates that the optimised implementation on the multiple GPU platform is a potential best solution for real-time aggregate risk analysis; the implementation is 77x faster than the sequential implementation on the CPU. 

\section{Conclusion and Future Work}
\label{conclusion}
In short, this paper has presented the aggregate risk analysis algorithm for analysis of insurance/reinsurance contracts, and its sequential and parallel implementations on multi-core CPUs and many-core GPUs. Large data is provided as input for aggregating risks across the Layers, and therefore, challenges such as efficiently organising input data in limited memory available, and defining the granularity at which parallelism can be applied to the aggregate risk analysis problem for achieving speed-up is considered. While the implementation of the algorithm on the multi-core CPU provides a speed-up of nearly 3x times over the sequential implementation, the basic GPU implementation provides a speed-up of approximately 9x times over the sequential implementation on the CPU. The most optimised implementation provides a speed-up of 16x times on the GPU and a speed up of \textbf{77x} on a multiple GPU over the CPU. It is notable that the acceleration has been achieved on relatively low cost and low maintenance hardware compared to large-scale clusters which are usually employed. These results confirm that high-performing aggregate risk analysis using large data for real-time pricing can be achieved on the GPU. 

Future work will aim to investigate the use of compressed representations of data in memory and to incorporate fine grain analysis, such as secondary uncertainty in the computations. 

\bibliographystyle{IEEEtran}
\bibliography{references}
\end{document}